\documentclass[useAMS,usenatbib,twocolumn]{mn2e}
\usepackage{subfigure}
\usepackage{rotating}
\usepackage{times}
\usepackage{graphicx}
\usepackage{lscape}

\newcommand{\ha}{H${\alpha}$}
\newcommand{\hb}{H${\beta}$}
\newcommand{\st}{[S\,{\sc II}]}
\newcommand{\nt}{[N\,{\sc II}]}
\newcommand{\zd}{[O\,{\sc III}]}
\newcommand{\zt}{[O\,{\sc II}]}
\newcommand{\nao}{Na\,{\sc I}}
\newcommand{\cat}{Ca\,{\sc II}}

\def\etal{{\it et al.}}

\def\sqiggt{\hbox{\rlap{\lower.55ex \hbox {$\sim$}}\kern-.05em \raise.4ex \hbox{$>$}\,}}
\def\sqiglt{\hbox{\rlap{\lower.55ex \hbox {$\sim$}}\kern-.05em \raise.4ex \hbox{$<$}\,}}

\begin{document}

\title[The short GRB~080905A]{Discovery of the afterglow and host galaxy of the low redshift short GRB 080905A\thanks{Based on observations at ESO telescopes at Paranal Observatory under program ID 081.D-0588}}

\author[A. Rowlinson \etal ]{A. Rowlinson$^{1}$\thanks{E-mail: bar7@star.le.ac.uk}, K. Wiersema$^{1}$, A. J. Levan$^{2}$, N. R. Tanvir$^{1}$, P. T. O'Brien$^{1}$,
\newauthor E. Rol$^{3}$, J. Hjorth$^{4}$, C. C. Th\"{o}ne$^{5}$, A. de Ugarte Postigo$^{5}$, J. P. U. Fynbo$^{4}$,
\newauthor P. Jakobsson$^{6}$, C. Pagani$^{1,7}$, M. Stamatikos$^{8,9}$\\ $^{1}$Department of Physics \& Astronomy,University of Leicester, University Road, Leicester, LE1 7RH, UK\\ $^{2}$Department of Physics, University of Warwick, Coventry CV4 7AL\\ $^{3}$Astronomical Institute `Anton Pannekoek'. University of Amsterdam, P.O.Box 94249, 1090 GE Amsterdam, the Netherlands  \\ $^{4}$ Dark Cosmology Centre, Niels Bohr Institute, University of Copenhagen, Juliane Maries Vej 30, 2100 Copenhagen, Denmark \\ $^{5}$ Istituto Nazionale di Astrofisica, Osservatorio Astronomico di Brera, via E. Bianchi 46, I - 23807 Merate \\ $^{6}$ Centre for Astrophysics and Cosmology, Science Institute, University of Iceland, Dunhagi 5, 107 Reykjav\'ik, Iceland \\$^{7}$ Department of Astronomy and Astrophysics, Pennsylvania State University, 525 Davey Laboratory, University Park, PA 16802 \\$^{8}$ NASA Goddard Space Flight Center, Greenbelt, MD 20771, USA \\$^{9}$ Center for Cosmology and Astro-Particle Physics (CCAPP) Fellow/Department of Physics, The Ohio State University, 191 West Woodruff Avenue,\\ Columbus, OH 43210, USA }


\date{Accepted 00. Received 00; in original form 00}

\pagerange{\pageref{firstpage}--\pageref{lastpage}} \pubyear{000}
\maketitle            

\label{firstpage}

\begin{abstract}
We present the discovery of short GRB 080905A, its optical afterglow and host galaxy. Initially discovered
by {\em Swift}, our deep optical observations enabled the identification of a faint optical afterglow, 
and subsequently a face-on spiral host galaxy underlying the GRB position, with a chance alignment probability
of $<$1\%. There is no supernova component present in the afterglow to deep limits. Spectroscopy of the galaxy provides a 
redshift of $z=0.1218$, the lowest redshift yet observed for a short GRB. The GRB lies offset from the 
host galaxy centre by $\sim 18.5$ kpc, in the northern spiral arm which exhibits an older stellar 
population than the southern arm. No emission lines are visible directly 
under the burst position, implying little ongoing star formation at the burst location. 
These properties would naturally be explained were the progenitor of GRB 080905A a 
compact binary merger. 
\end{abstract}

\begin{keywords}

Gamma-ray burst: individual: GRB 080905A

\end{keywords}

\section{Introduction}
The detection of the first fading afterglow from long gamma-ray bursts (LGRBs) \citep{costa1997, paradijs1997} 
proved to be a pivotal moment in their study. Similarly, the first identifications of short GRB (SGRB) 
afterglows in 2005 opened a new window on this still enigmatic class of transient \citep{gehrels2005, hjorth2005,
fox2005, berger2005}. Afterglows provide precise positions, and hence a 
route to redshifts and identifying the host galaxies. This in turn provides luminosities and space
densities for the bursts, allows possible means of measuring their collimation, and ultimately may 
enable ``smoking" gun signatures of their progenitors to be uncovered, as was the case for LGRBs 
\citep{hjorth2003}.

GRBs can be categorised, largely
according to their duration, as LGRBs and SGRBs, with the dividing line being 
at roughly 2s \citep{kouveliotou1993}. LGRBs have been identified with starforming 
galaxies with moderately low metallicity \citep[e.g.][]{bloom2002, fruchter2006} leading to the 
suggestion that young massive stars are the most likely progenitors \citep[collapsars, 
summarised in ][]{woosley2006}. Studies of the locations of LGRBs with the 
{\it Hubble Space Telescope} ({\it HST}) have shown that LGRBs typically occur in compact galaxies 
\citep{fruchter2006,wainwright2007}, with mean effective radii of 1.7 kpc, and in
the brightest regions of their hosts \citep{bloom2002,fruchter2006}. 
However, studies of SGRB host galaxies contrast with this picture. Some hosts contain
exclusively ancient populations \citep{berger2005,gehrels2005,bloom2006}, while others are actively
star forming \citep{fox2005,levan2006}, or contain a mixture of young and older populations \citep{soderberg2006}. 
Additionally, the bursts themselves can be significantly offset from their hosts 
\citep{berger2005, fox2005, bloom2006, troja}, a fact which can complicate the task
of host identification \citep{levan2007}. These observations imply that the
progenitors of SGRBs can originate from ancient populations, which may well
be kicked from their birthplace. The final mergers of 
compact object binaries remain a prime candidate \citep{lattimer1976, eichler1989, narayan1992}. 
Additionally, there is 
thought to be a separate population of SGRBs at redshifts $\ll 0.1$ which are associated with 
giant flares from extra-galactic Soft Gamma Repeaters \citep{hurley2005, tanvir2005}. At very 
low redshift there may be overlap between these, and a more energetic, cosmological population.  

With the influx of GRBs over the past few years, it has become apparent that there are examples of 
SGRBs and LGRBs which have properties which make it hard to unambiguously assign them to one
category or the other. For example GRB 050724, $T_{90}=3 s$, occured in a nearby elliptical galaxy 
which, among other properties, associated it with the SGRB population \citep{barthelmy2005, berger2005}. 
GRB 060505, $T_{90}=5 s$, had no associated supernova but had a spectral lag which is 
consistent with the LGRB population, was located within a star formation region in a host galaxy at 
$z=0.09$ and its classification is still not firmly established \citep{noSN,ofek2007,jakobsson2007,Thoene,
mcbreen2008,bloom2008,xu2009}. Additionally, GRB 060614, 
$T_{90}=100 s$, is thought to be a SGRB with extended emission as there was no associated supernova and 
the spectral lag was consistent with other SGRBs \citep{noSN,galyam2006,zhang2007}. Recently, GRB 090426 was identified 
as the most distant SGRB at a redshift of $z=2.609$ and $T_{90}=1.28s$, \citep[][Th\"one et al. In Prep]{levesque2009}. 
\cite{levesque2009, antonelli2009} conclude from their 
observations that the simplest explanation is that the progenitor was more likely to be a collapsar. With 
this in mind, it is imperative that we understand the properties of  host galaxies and, if possible, 
the local environment of GRBs to aid in their classification and the identification of the progenitor.

The majority of GRBs lie at moderate to high redshift, and have host galaxies which
are at best only marginally resolved from the ground. Most SGRB redshifts to date come from the putative host 
galaxy and in some cases these may be ambiguous \citep{levan2007}. GRBs 060505 \citep{Thoene}, 020819 \citep{levesque2010} 
and 980425 \citep{michalowski2009} have been identified in nearby host galaxies, which has allowed more detailed 
spectroscopy. Instead of relying on the properties of the galaxy as a whole, it has been possible to 
subdivide the galaxy into relevant regions and complete spatially resolved spectroscopy. This has 
allowed the study of the region in which the GRB occured giving details about the local stellar 
population. 

Here, we present the discovery of the optical afterglow, and host galaxy of the short GRB 080905A. 
Its faint afterglow pinpointed its location to a spiral host galaxy at $z=0.1218$, the most local 
short burst yet known. SGRB 050709 has the next lowest confirmed redshift for a SGRB at $z=0.16$ 
\citep{fox2005}, followed by SGRB 050724 associated with a host galaxy at $z=0.257$ 
\citep{barthelmy2005, berger2005}. SGRB 061201 may be associated with a galaxy at lower redshift 
of $z=0.111$ but it was not possible to confirm this as it was offset by 17" \citep{stratta2007}. 
In section 2, we describe the observations obtained of the afterglow of GRB 080905A and the 
spectra obtained for the host galaxy. We  analyse these data in section 3, discuss the implications 
of our findings in section 4 and draw our conclusions in section 5.

Throughout the paper we adopt a cosmology with $H_0 =  71$ km\,s$^{-1}$\,Mpc$^{-1}$,  
$\Omega_m = 0.27$, $\Omega_\Lambda = 0.73$. A redshift of $z = 0.1218$  then gives a luminosity 
distance of 562.3 Mpc, and 1 arcsecond corresponds to 2.17 kpc. The errors are quoted at 90\%
confidence for X-ray and 1$\sigma$ for optical.

\section{Observations and Analysis}

\subsection{Prompt emission properties}
GRB 080905A was detected by Swift at 11:58:54 UT. It is a SGRB with T$_{90}$ duration of 1.0 
$\pm$ 0.1 s, the Burst Alert Telescope (BAT) detected three flares peaking at 
$T+0.04067\pm0.0007$ s, $T+0.17^{+0.03}_{-0.10}$ s and $T+0.869\pm0.003$ s. The time averaged BAT 
spectrum was best fit by a power law with a photon index of $\Gamma=0.85\pm0.24$ and the fluence was 
(1.4 $\pm$ 0.2) $\times$ 10$^{-7}$ erg~cm$^{-2}$ in the 15 - 150 keV energy band \citep{cummings2008}. The flux at 
a specific frequency, $\nu$, and time is given by $f_{\nu}\propto\nu^{-\beta}t^{-\alpha}$ where $\beta=\Gamma-1$. GRB 
080905A was also detected by {\it INTEGRAL} \citep{pagani2008} and {\it Fermi} Gamma-ray Burst Monitor (GBM) 
\citep{bissaldi2008}. Using the redshift of 0.1218, the isotropic energy released is $4.7\pm0.7\times10^{49}$ erg in 
the 15 - 150 keV energy band.

Some short GRBs show evidence for a soft extended emission component in the prompt 
emission \citep[e.g][]{barthelmy2005, norris2006}. There is no evidence of soft extended emission 
in the BAT 15-25 keV light curve for GRB 080905A, with a limiting flux of $<5.2 \times 10^{-7}$ erg 
cm$^{-2}$ s$^{-1}$.

Additionally, short GRBs have negligible spectral lag in their prompt emission unlike long GRBs 
\citep{norris2006,yi2006}. We performed a spectral lag analsysis of GRB 080905A based upon the cross correlation
function methodology used in \cite{ukwatta2010}. The analysis considered several timescales using 128, 64, 32, 
16, 8 and 4 ms binned lightcurves and compared all six pairing combinations of the BAT's four energy 
channels. There is a lack of emission below 25 keV, which results in very low cross correlation amplitudes 
for paired lightcurves containing channel 1. Channels 2 and 4 also have relatively low emission, so in 
this analysis we use cross correlation between channels 2 and 3 and determine the lag using a gausian fit. We 
calculate the 1$\sigma$ error using 1000 lag Monte Carlo simulations. The lag time of GRB 080905A is $4\pm17$ 
ms, which is consistent with zero as expected for a short GRB.

\subsection{X-ray Afterglow Observations}

The fading X-ray afterglow was located by {\it Swift} with an enhanced position of RA (J2000): 19 10 41.74 and Dec (J2000): -18 52 48.8 with an 
uncertainty of 1.6 arcsec (90\% confidence) \citep{evans2008b}.

The time averaged X-ray Telescope (XRT) spectrum using the photon counting (PC) data is best fit by an 
absorbed power law with photon index $\Gamma=1.45\pm0.25$ and with an intrinsic absorption $N_{\rm H} = 1.6\pm1.0\times10^{21}$ 
cm$^{-2}$ in excess of the Galactic absorption of $N_{\rm H} = 9\times10^{20}$ cm$^{-2}$ \citep{kalberla2005}. 
We fit the combined BAT/XRT lightcurve with a power law decay models with one break. The best fit spectrum is 
$\alpha_{1} = 2.62^{+0.25}_{-0.13}$, breaking at $T_{1}= 443^{+408}_{-84}$ s to a decay of 
$\alpha_{2}=1.49^{+0.60}_{-3.66}$. The BAT count rates 
(in the 15-150 keV energy band) were extrapolated to the XRT energy band (0.3-10 keV) and converted to flux 
using the average spectral index for the BAT and XRT PC spectra and standard tasks in {\sc Xspec}. These fluxes 
were then combined with the XRT lightcurve to create the combined BAT/XRT lightcurve. Using the redshift of 0.1218 
and a k-correction \citep{bloom2001}, the combined BAT/XRT lightcurve has been converted to the rest-frame time and 
luminosity and is shown in Figure 1.

\begin{figure}
\centering
\includegraphics[width=8.2cm]{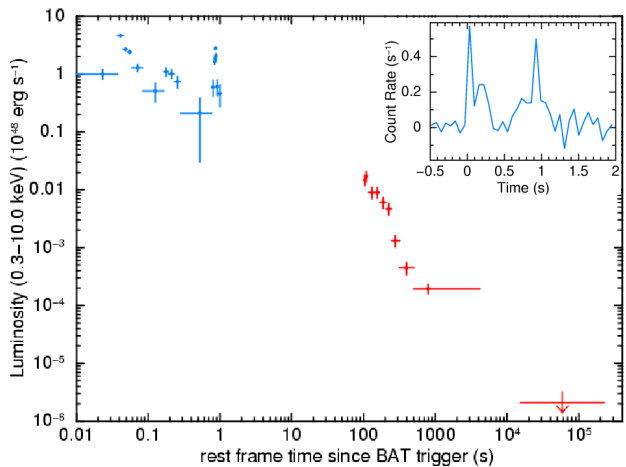}
\caption{This shows the combined BAT and XRT luminosity and rest frame light curve for GRB 080905A. The BAT 
data are plotted until $\sim2 s$ and the XRT data are plotted starting at $\sim100s$. Inset is the BAT 
lightcurve with linear observed time on the horizontal axis and BAT count rate on the vertical axis.}
\end{figure}

\begin{figure}
\centering
\includegraphics[width=8.2cm]{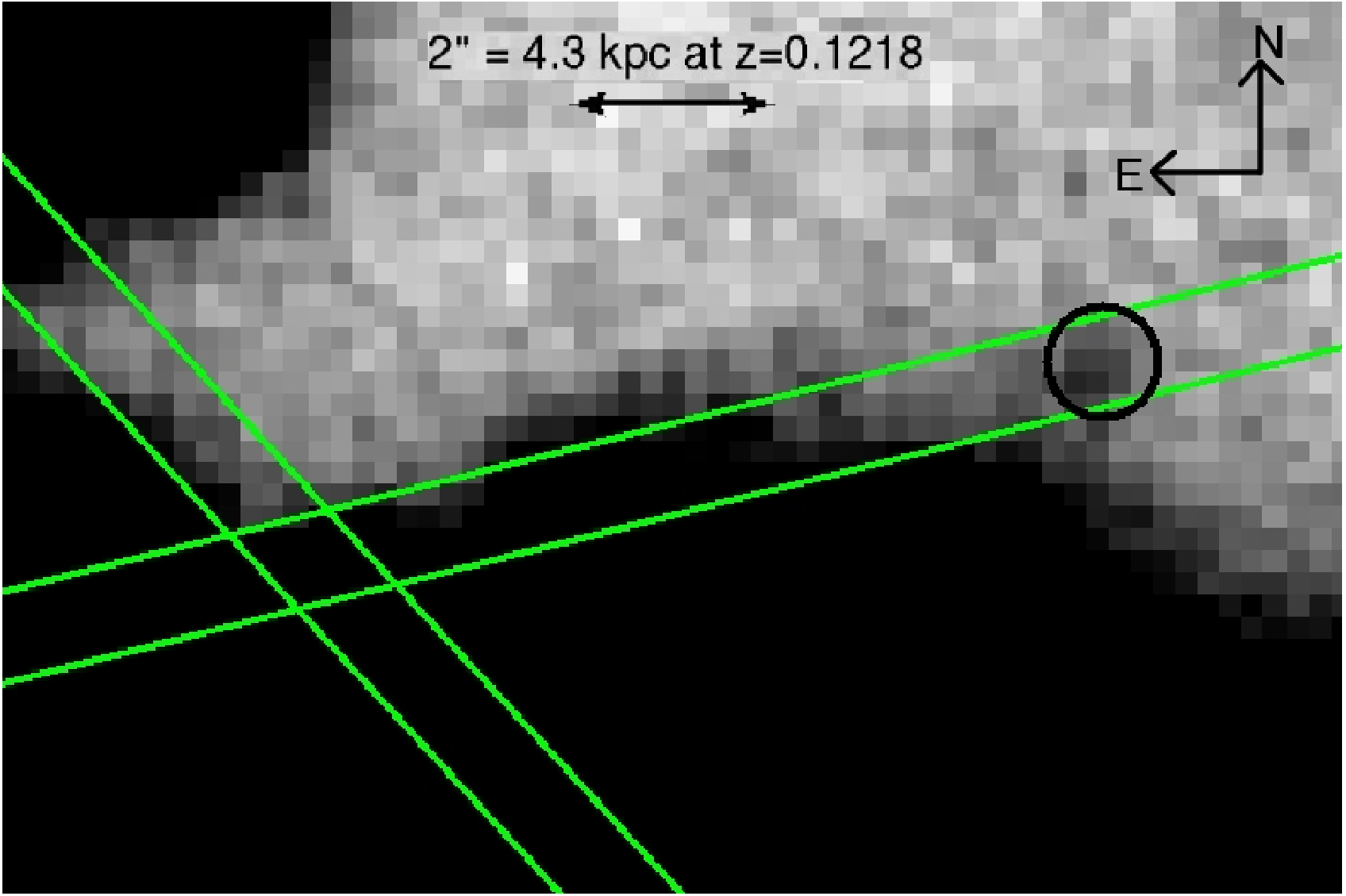}
\includegraphics[width=8.2cm]{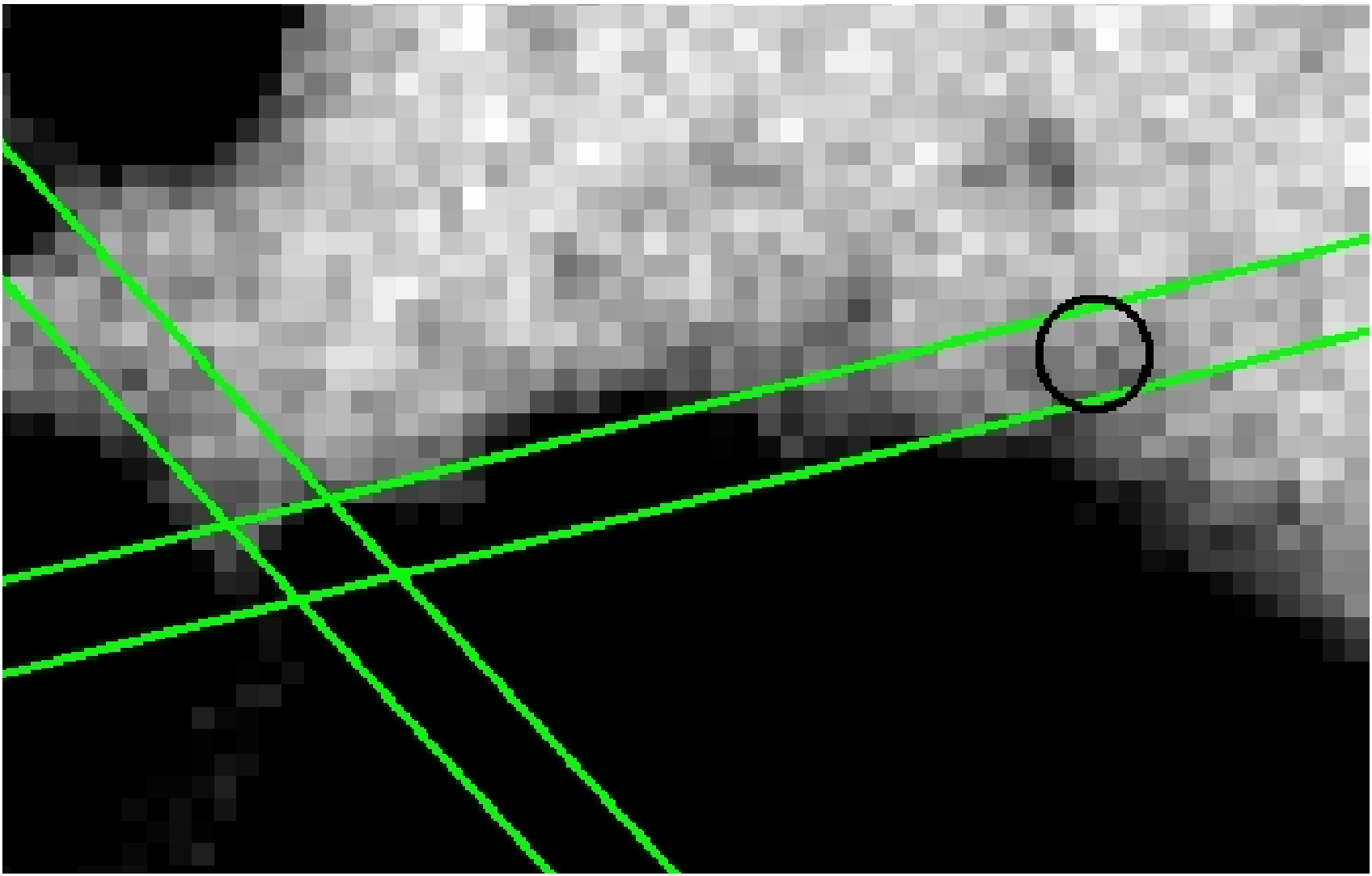}
\caption{The circle marks the location of the afterglow of GRB 080905A, the top image is from epoch 2, 
observed 14.3 hours after the trigger time, and the bottom image is from epoch 4, $\sim$18 days after 
the trigger time (see Table 1). For reference, the two slit positions used for spectroscopy have 
also been included.}
\end{figure}

\begin{figure}
\centering
\includegraphics[width=8.2cm]{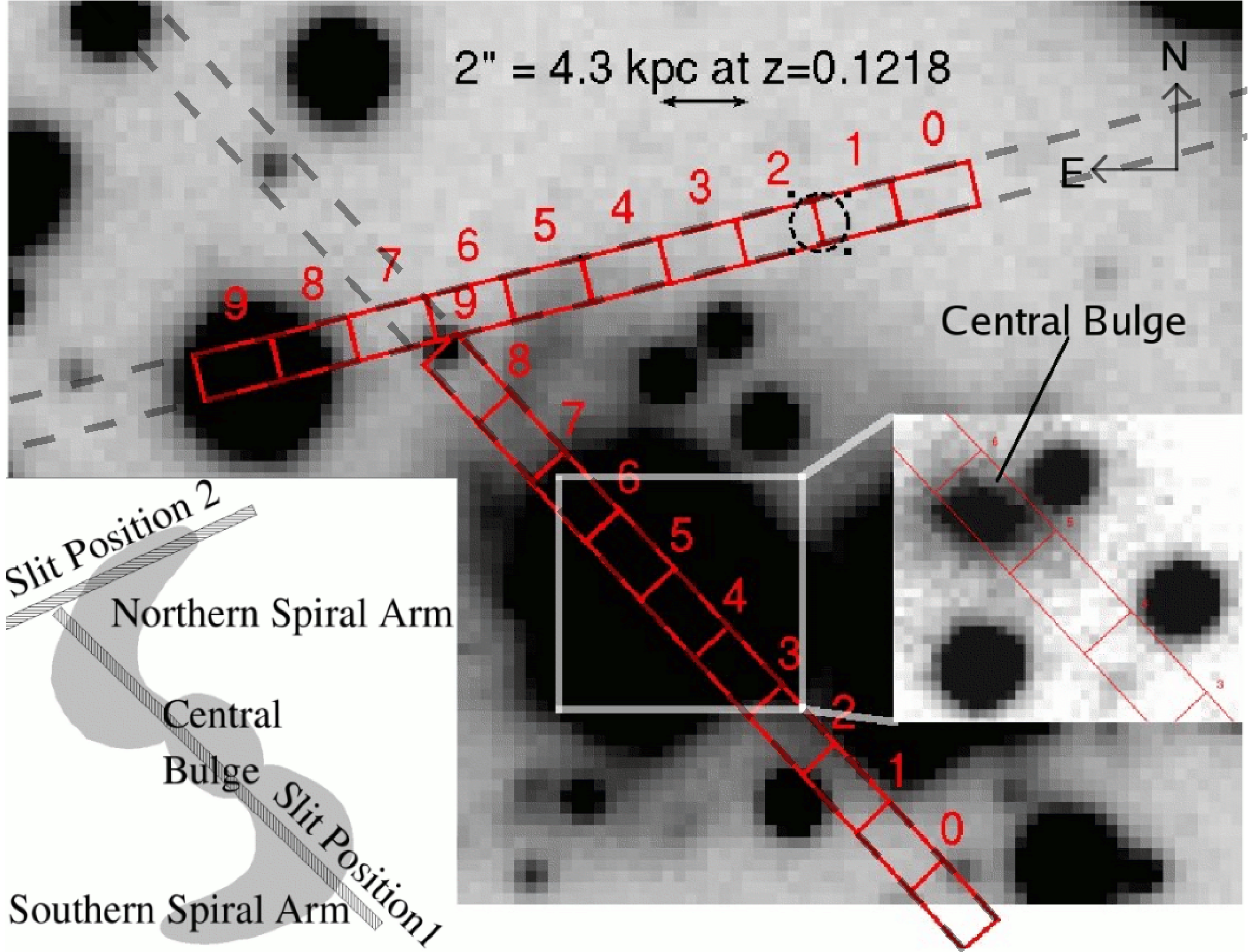}
\includegraphics[width=8.2cm]{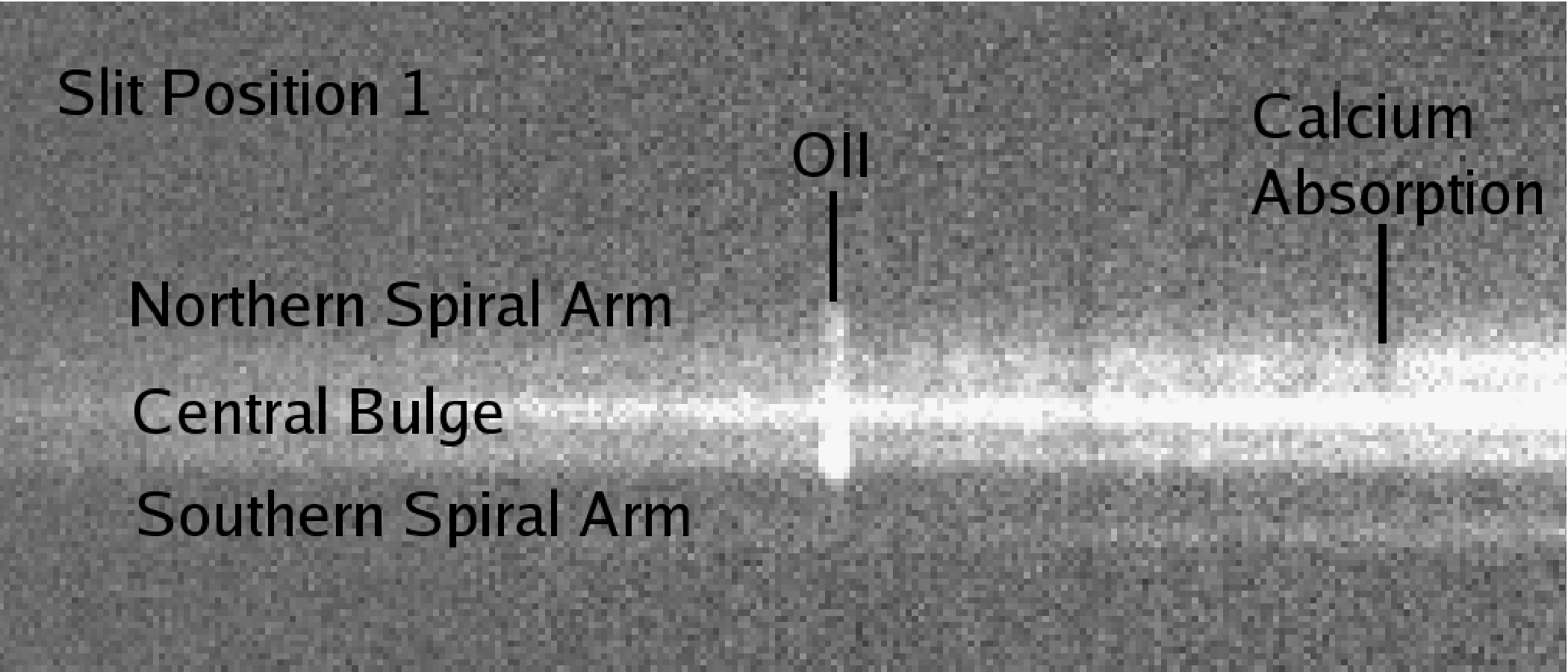}
\includegraphics[width=8.2cm]{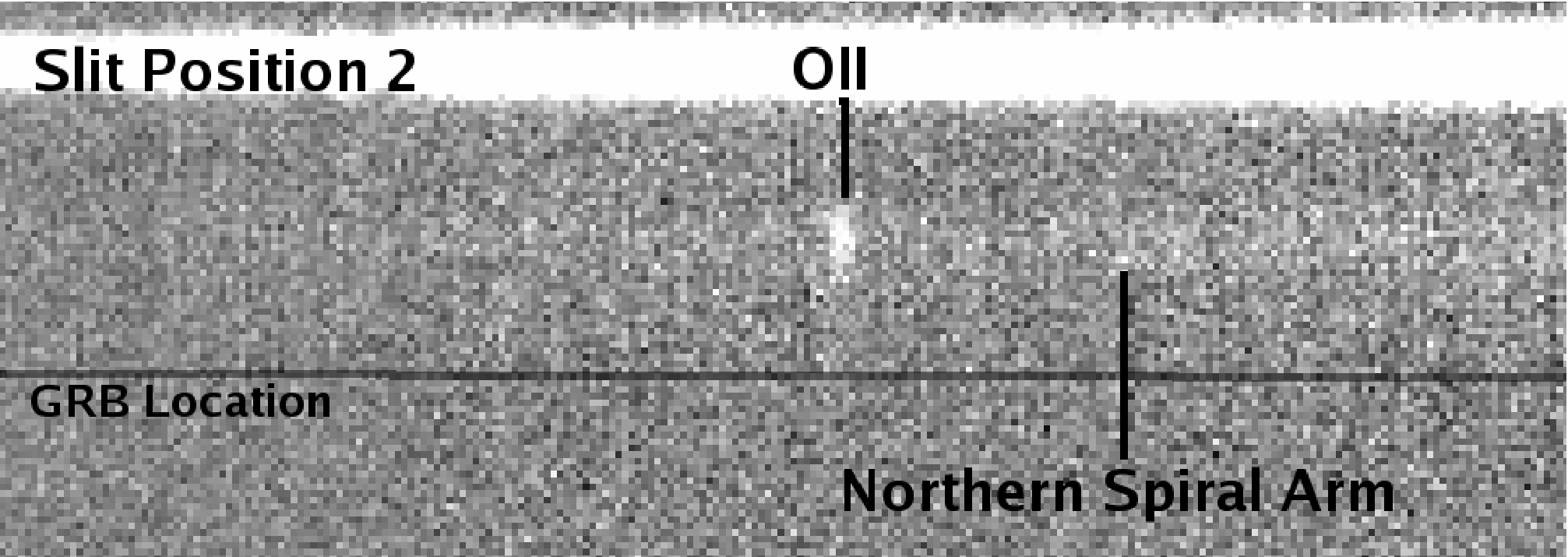}
\caption{This shows the locations of the two slit positions used to obtain the spectra (dashed lines) and the subapertures 
into which the spectra was divided into subspectra. The dashed circle shows the location of the optical 
afterglow. The main image shows the spiral arms in the R band. The inset on the right shows the 
central bulge of the galaxy in the K band. In the bottom left corner, there is a sketch of the 
structure of the galaxy. In the middle panel is the 2D spectra for slit position 1 in which the emission 
lines get fainter when moving from the southern spiral arm to the northern 
spiral arm and the absorbtion features dominate more in the northern spiral arm than in the 
southern spiral arm. Additionally, the continuum in the northern spiral arm is fainter than the 
southern spiral arm bluewards of the D4000 break. In the bottom panel is the 2D spectra for slit position 2 in 
which faint emission lines can be observed and the continuum of the northern spiral arm. The horizontal line shows the
location of the GRB.}
\end{figure}

\subsection{Optical Observations}
Early optical imaging of GRB 080905A obtained only upper limits on the afterglow flux, which 
were found by UVOT at T+ 114 s (V $>$21.3 \citep[][]{brown2008}), 
the Mt. John Observatory at T+ 2580 s (R $>$20.8 \citep{tristram2008}), and the MITSuME 
telescope at T+ 2520 s (R $>$17.6) \citep{nakajima2008}. 

Our observations began at the Nordic Optical Telescope (NOT) 8.5 hours after the burst, with further epochs
obtained with the Very Large Telescope (VLT) utilizing FORS2 taking place 14.3 and 36 hours after the burst. A final 
R-band observation was made on 23 September, 17.5 days post burst. Using ISAAC we obtained a further K-band observation 
on 1 October, 25.5 days post burst. Comparison of these observations
allowed us to discover both a faint optical afterglow, and an underlying spiral host galaxy 
\citep{malesani2008,postigo2008}. 

Our optical images were reduced in the standard fashion, and magnitudes for the afterglow 
derived in comparison to USNO and 2MASS objects within the field (since conditions were not photometric
at the time of the observations). As the afterglow lies on the edge of its spiral host we obtain
host subtracted afterglow fluxes by subtraction of the light from this galaxy, assuming zero
contribution of transient light in the final epoch of optical images. The resulting magnitudes are shown in 
Table 1. The afterglow is faint $R \sim 24$, even for a SGRB, and demonstrates
the necessity of deep and rapid observations in the location of SGRBs. Converting the optical magnitude 
of GRB 080905A to a flux of $\sim 7\times10^{-30}$erg cm$^{-2}$ s$^{-1}$ Hz$^{-1}$ and comparing it to 
the sample at 11 hours considered by \cite{nysewander2009}, it is one of the faintest afterglows detected 
and, with a optical luminosity of $\sim 6.7\times10^{25}$erg s$^{-1}$ hz$^{-1}$, the lowest luminosity optical 
afterglow detected and lies below the trend observed between optical afterglow intensity and isotropic 
energy, suggesting that this GRB occured in a low density environment. We used a reasonable extrapolation of the X-ray 
light curve to the time of our optical imaging to determine that the non-detection of the X-ray 
afterglow is consistent with the decay observed. The location of the optical afterglow is RA(J2000): 
19 10 41.71 and Dec(J2000): -18 52 47.62, with an error of 0.76 arcseconds, and is shown in Figure 2.

The afterglow is located $\sim$9" from the centre of an R$\sim$18 galaxy and we conclude that this is the 
host galaxy. To calculate the liklihood of a chance alignment of a similar or brighter galaxy within 10" 
of the afterglow we use the the host galaxy magnitude and size and the number of galaxies of this magnitude 
or brighter \citep{hogg1997}. The probability of a chance alignment is less than 1\%. A more accurate 
method would be to use the half light radius of the galaxy as described in \cite{fong2009}, however 
it is difficult to calculate this due 
to contamination of foreground stars. The low chance probability and that the afterglow location lies 
within the stellar field of the galaxy, both support our conclusion that this is the host galaxy of GRB 
080905A. As for many GRBs without afterglow redshifts, it is possible that GRB 080905A is associated
with a higher redshift galaxy which is fainter than the deep limiting magnitude of our optical images (R$>$25).

The location of the afterglow is offset from the centre of the host galaxy by a projected radial distance of 
18.5 kpc. This is a relatively large offset, but is comparable to several other SGRB locations 
\citep{troja,fong2009} and it is important to note that the host galaxy is relatively large so the 
host-normalised offset would be much smaller. Host-normalised offsets are calculated by normalising 
the offset to the effective (half light) radius of the host galaxy.

At a redshift of $z=0.1218$ a supernova like SN~1998bw would reach a peak magnitude of
roughly $R \sim 19.5$, a factor of $>$100 brighter than any object present in our final epoch. 
The lack of any visible supernova component is in keeping with searches which have been done in other 
SGRBs, supporting our classification of GRB 080905A as a member of the SGRB population. For example, \cite{hjorth2005} 
conducted an early search for a SN component for SGRB 050509B finding any accompanying SN would 
be fainter than typical SNe and \cite{fox2005} conducted place deep 
limits for a SN component for SGRB 050709. Our observations can also be used to probe the possible 
production of radioactive Nickel during GRB 080905A. The ejection of radioactive material in the process 
of an NS-NS merger may create a visible electromagnetic signal described as a mini-SN 
\citep{li1998,kulkarni2005,metzger2008,kocevski2009}. The absence of any late time emission brighter than $R \sim 24 $, 
coupled with the known low redshift makes these constraints strong in the case of GRB 080905A, although 
the cadence of the observations is sensitive to either relatively fast, or slow rise time (but not those of
intermediate duration). This suggests than the radioactive yield associated with GRB 080905A is 
$<0.01 M_{\odot}$, based on the low redshift model developed by \cite{perley2009} for GRB 080503 and the 
general models in \cite{kulkarni2005}.

\begin{table*}
\begin{tabular}{@{}cccccccccc@{}}
\hline
Epoch & Date mid point & Time after trigger & Telescope & Exposure time & Filter/grism & Seeing  & Magnitude\\ 
& (UT) & (hours) & & (s) & & (arcsec) &  \\
\hline
1 & Sep 05 20:30 UT & 8.5 & NOT &  1800 & R & 0.9 &   $24.04 \pm 0.47$\\
2 & Sep 06 02:39 UT & 14.3 & VLT & 2400 & R special & 1.05 &  $24.26 \pm 0.31$\\
3 & Sep 07 00:29 UT & 36 & VLT & 2400 & R special & 0.85 &  $>25.0$\\
4 & Sep 23 00:44 UT & - & VLT & 2400 & R special & 0.85 &  -\\
5 & Oct 01 01:15 UT & - & VLT & 7200 & K short & 0.65 &  - \\
\hline
  (slit position 1) & Sep 24 01:39 UT & - & VLT & 3600 & grism & 0.9 &  - \\   
  (slit position 2) & Sep 24 02:27 UT & - & VLT &  3600 & grism & 0.9 & - \\
\hline
\end{tabular}
\caption{Log of observations of the afterglow and host of GRB 080905A. The magnitudes shown for the afterglow are
host subtracted, assuming zero contamination from the afterglow in epoch 4. Magnitudes have been
corrected for foreground extinction of $E(B-V) = 0.14$.   \label{table:log}}
\end{table*}

\subsection{Host Galaxy Spectroscopy}

To characterize the host galaxy, we obtained deep spectroscopy on  September 24th 2008, 
using FORS1 on UT2 of the VLT, Chile. These observations were obtained after the optical
afterglow had faded. To maximize wavelength coverage we used the 300V grism with 
the GG375 filter to supress contamination by the second spectral order. This results in a 
wavelength range $\sim$3700 to 9200 \AA.
 
The 1.0 arcsecond wide slit was oriented along two different fixed position angles, illustrated in 
Figure 3, and $4 \times 450$ second exposures were acquired for each slit position. The two slit 
positions (-104.1 and -42.7 degrees) were chosen to cut through the host galaxy covering 
the nucleus as well as spiral arms on either side of the galaxy (hereafter ``slit position 1''), and to 
cover the afterglow position and cut through a nearby spiral arm (hereafter ``slit position 2''). 
Seeing conditions during the observations were reasonable with an average seeing of 0.9 arcseconds and
mean airmass of 1.2 (slit position 1) and 1.3 (slit position 2). 
We reduced the data using standard procedures in IRAF. The four exposures per slit position were combined 
before extraction, removing cosmic rays in the process.

We extracted the spectra of slit position 1 and 2 in the same way: we use the relatively bright continuum of 
the bulge (slit position 1) or a nearby bright star (slit position 2) to fit the shape of the trace function, and 
extract using 10 adjoining, equally sized subapertures following this trace. Subapertures are 7 
pixels in size in both slit position 1 and 2 data, which corresponds to 1.76 arcseconds per subaperture 
(pixel scale is 0.252 arcseconds per pixel), i.e. a value matched to twice the seeing full-width-at-half-maximum
(FWHM). At the redshift of the host galaxy, this corresponds to a physical scale of 3.8 kpc per subaperture.
In the following we will refer to the spectra extracted with these small apertures as subspectra. Figure 4 shows 
examples of extracted subspectra. The GRB location is covered only by slit position 2, and falls in subapertures 
1 and 2. The subspectra are wavelength calibrated using He, HgCd and Ar lamp spectra. From the FWHM of a Gaussian 
fit on the arc lines we measure a nominal spectral resolution of 11 \AA\ at the central wavelength.    

Flux calibration of the subspectra was done using observations of the spectrophotometric standard star 
LDS 749B, Atmospheric extinction correction was done by applying the average CTIO atmospheric 
extinction curve. A Galactic dust extinction correction was performed by using the $E(B-V)$ value of
 0.14 \citep{Schlegel}, assuming a Galactic extinction law $A_{\lambda}/A_{V}$ expressed as 
$R_{V} = A_{V}/E(B-V)$ \citep{Cardelli}. We make the standard assumption $R_{V} = 3.1$ \citep{Rieke}.
No Galactic \nao\ or K\,{\sc I} absorption is detected in the spectrum, consistent with the 
$E(B-V)$ value from \cite{Schlegel}. Note that this calibration provides us with a good 
{\em relative} flux calibration needed to evaluate changes between the different subspectra, e.g in 
the strength of emission line ratios or some continuum features, but does not provide a full 
absolute calibration. 
  
From the detected emission lines we measure the redshift of the GRB host galaxy to be $z = 0.1218 \pm 
0.0003$.  

\section{Host galaxy properties \label{sec:hostg}}

\subsection{Host morphology \label{sec:morphology}}

Visual inspection of the images in R and K bands shows the host to be a nearly face-on galaxy, with
clear bulge, disk and spiral arm components. At least two spiral arms can be distinguished, one 
on either side of the galaxy, which are hard to see due to a great number of foreground stars. 
Figure 3 shows the spiral arms as observed in the R band and inset is an image of the central bulge 
in the K band. The detection of spiral arms in combination with the detected emission 
and absorption lines allows us to loosely classify the host of GRB 080905A as an Sb/c galaxy.

By subtracting foreground stars from the images we estimate that the host of GRB 080905A
has an R-band magnitude of R=$18.0 \pm 0.5$. The large error arises not due to the faintness
of the object, but due to the uncertainty in subtracting the significant number of foreground sources which
overlap the spiral structure. Correcting for foreground extinction ($E(B-V) =0.14$)
this corresponds to an absolute magnitude of $M_V \sim -21$, and
suggests that the host of GRB 080905A is broadly similar to the Milky Way. 

Using the near-infrared mass-light ratio, $M_{*,old}(M_{\odot})=2.6 \times 10^{8} D^{2}(Mpc)F_{k}(Jy)$ 
\citep{thronson1988}, and the K-band magnitude of the host galaxy, $K=16.2^{+0.4}_{-0.7}$, we determine 
the mass of the host galaxy to be $M_{*,old}=2 \pm 1 \times 10^{10} M_{\odot}$. The errors are estimated 
based on the uncertainties in subtraction of the foreground stars and identifying the extent of the host galaxy.

\subsection{Rotation curve \label{sec:rv}}

The nearly face-on orientation of the host galaxy gives us an excellent view of the location of 
the GRB within the host, similar to GRB 060505 which also occured in nearly face-on 
Sbc galaxy \citep{Thoene}. However, this favourable geometry complicates measurements of the host dynamical 
mass, required to test the consistency of this host, and GRB spiral hosts in general, with the 
mass -- metallicity relation at this redshift.

Visual inspection of the 2D spectrum shows no clear slant in the \st, \ha, \nt, \zd, \hb\ and 
\zt\ lines (\zt\ are shown in Figure 4). To determine the rotation curve of the galaxy (or upper limits) we  
use the {\it fxcor} routines in the IRAF {\it rv} package to Fourier cross-correlate the spectra 
of different subapertures of the slit position 1 spectra, finding their relative radial velocity as a 
function of distance to the galactic nucleus. We correlate spectral sections around the brightest 
emission lines, as well as the full subaperture spectra (using also absorption features). 
We fit a Gaussian function to the cross correlation peak to determine its centre and width. Between the 
two subspectra with the highest signal to noise emission lines, subapertures 3 and 7, 
we find a formal radial velocity difference of $19 \pm 38$ km\,s$^{-1}$.  Using symmetrical 
subapertures about the galactic centre (4 and 8) we find 30 +/- 160 km\,s$^{-1}$. This value is using 
very weak emission lines in subaperture 8, so is a much less constraining limit.

Using the GALFIT software package \citep{galfit} we decompose the host galaxy to identify the 
inclination angle. We use the acquisition images for the spectra, which have the best seeing 
conditions. We use an empirical PSF as modelled through the IRAF {\sc DAOPHOT} routines 
using several moderately bright stars close to the GRB position. We find an inclination angle of 
$\sim23^{\circ}$, however there are large errors associated with this value due to poor signal 
to noise, contamination by bright stars and the near face-on inclination. This angle appears to be smaller 
than that identified for LGRB 980425, $\sim50^{\circ}$ \citep{christensen2008}.

\subsection{Spatially resolved properties}

The middle and bottom panels of Figure 3 show the subspectra from slit position 1 and 2, in which differences in continuum 
shape and line properties can be seen, reflecting subtle changes in stellar population properties 
dominating the differing subspectra. From Figure 3 it is clear that several field stars are 
located close to and on top of the host. Some of the subspectra appear affected by light from 
these stars, which can be seen by the presence of Balmer, Na and Ca lines at zero redshift, and 
from the shape of the continuum. 

\begin{figure*}
\centering
\includegraphics[width=16.4cm]{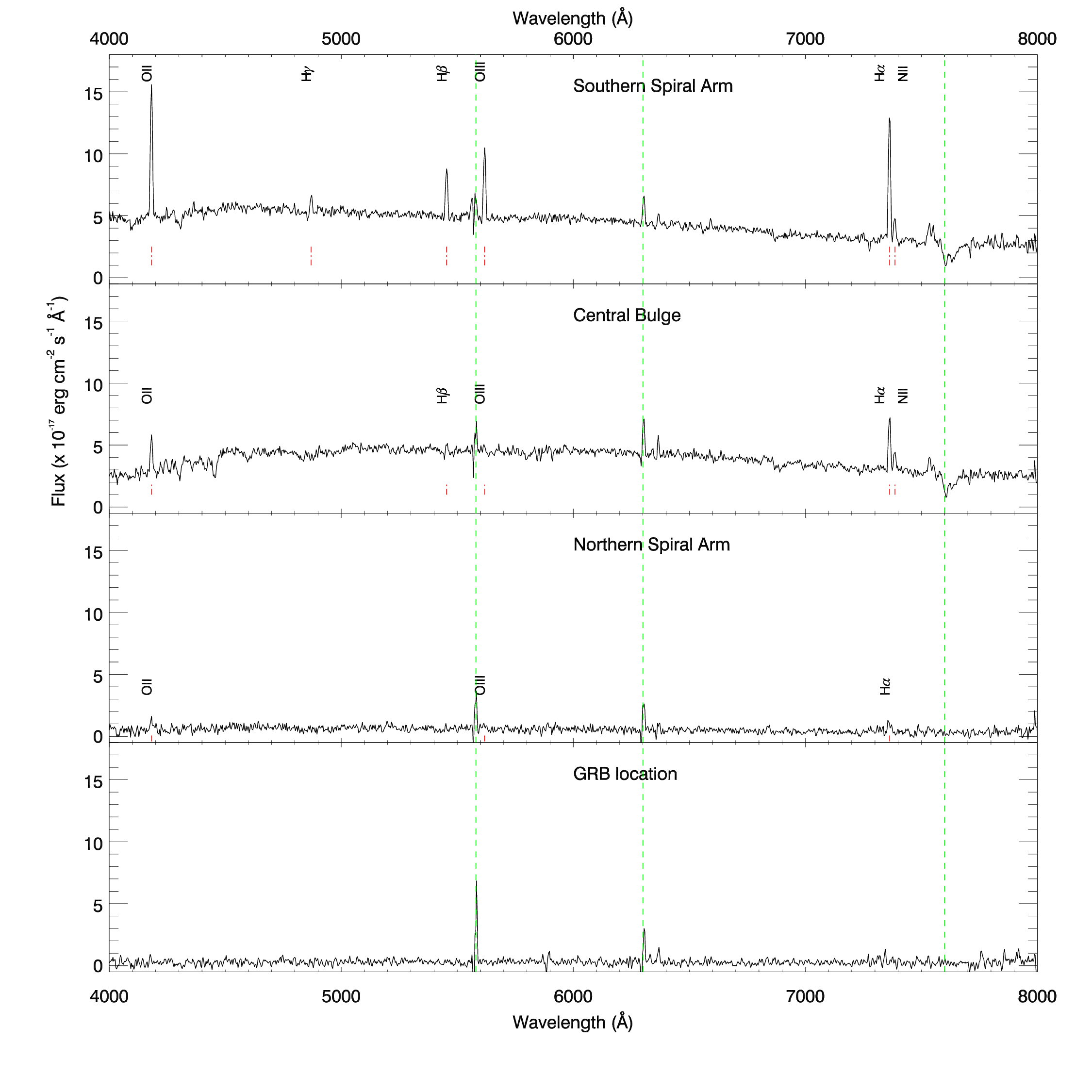}
\caption{This shows the observed spectra in the northern spiral arm, the central bulge and the southern spiral arm. These 
correspond to subapertures 4, 6 and 8 from slit position 1. The lowest panel shows the observed 
spectrum at the GRB location. There are residual features from sky line subtraction at $\sim$5600\AA, 
$\sim$6300\AA\ and $\sim$7600\AA\ (shown by the dashed lines).}
\end{figure*}

The 2D spectra show clearly several basic properties of the host. In Figure 4, we show subspectra 
from subapertures 4 (northern spiral arm), 6 (central bulge) and 8 (southern spiral arm) from slit 
position 1. Additionally, we show the subspectra from subaperture 2 for slit position 2, corresponding 
to the GRB location. The slit position 
1 spectrum shows that the nebular emission lines are strongest in the southern part of the host, 
and get dramatically weaker northwards of the nucleus. This shows that the star formation rate is 
strongest in the spiral arm diametrically opposite the GRB position, in stark contrast to the 
spiral host galaxy of LGRBs 980425, that show strongest star formation at, or near, 
the location of the burst \citep{christensen2008}. The GRB location appears to lie in the extension 
of a spiral arm. The 2D spectrum of slit position 2, which probes this arm, clearly shows strong nebular 
emission lines of \zd\ and \zt, and weaker \ha\ and \hb\ at the location of the spiral arm, but no 
emission line flux is detected at the location of the burst. At and near the GRB location a weak, 
near featureless continuum can be seen.

The slit position 1 subspectra that are dominated by bulge light show clear absorption features
common to old populations and ISM gas (\nao, \cat, 4000 \AA\ break, G band), and show stellar 
atmosphere Balmer absorption underneath the nebular Balmer emission. The other spectra have 
brighter nebular lines and weaker 4000 \AA\ breaks. As several of the subspectra are contaminated 
by light from foreground stars, and the resolution of the spectra is low, we limit our analysis 
in this paper to the nebular emission lines and the strongest absorption bands. 

Using the relative fluxes of H$\alpha$ and H$\beta$, we are able to determine the flux ratio at different 
points in the host galaxy, as shown in Figure 5. This gives an indication of the reddening in the host 
galaxy, which is important to consider as the metallicties and D4000 we calculate may be 
affected by this value. Figure 5 shows that the southern spiral arm and central bulge are consistent 
with having little significant reddening. However, the northern spiral arm shows significant reddening, 
and this will affect our R23 calculations.

We measure the emission line fluxes in each subspectrum in slit position 1, and compute the metallicity 
profile along this slit position through the $N2$ indices, where $N2\equiv[{\rm N_{II}}]\lambda6583/{\rm H}\alpha$ \citep{pettinipagel}.   
In addition to these indices, we compute R23 metallicities where possible using $R23\equiv([{\rm O_{II}}]+[{\rm O_{III}}])/{\rm H}\beta $ \citep{pettinipagel}. 

In the slit position 2 subspectra we can only determine emission line upper limits at the GRB location. 
The metallicity of the spiral arm that is covered by slit position 2 can be calculated through R23.

\begin{figure*}
\centering
\includegraphics[width=16.4cm]{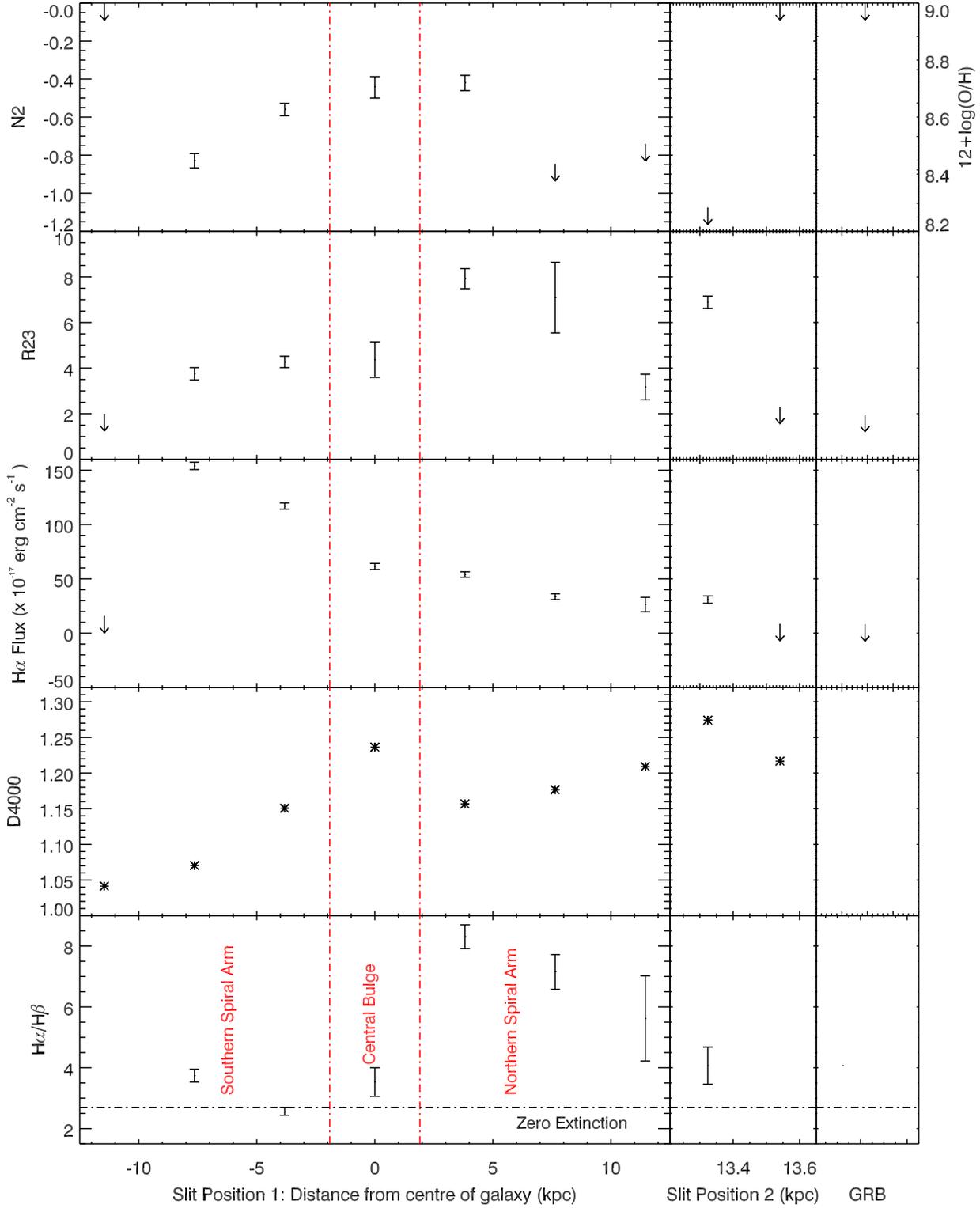}
\caption{Panel 1 shows the Log(N2), H$\alpha$ Flux, R23, D4000 and the flux ratio of H$\alpha$ to H$\beta$ 
as a function of position from the centre of the host galaxy. Note that the H$\alpha$ Flux can only be used as a relative value 
as it has not been absolutely calibrated. In the lowest panels, we show the H$\alpha$ to H$\beta$ ratio for zero extinction. 
In the second panel is the data from slit position 2 along the spiral arm and the third shows the upper limits for the region 
in which the GRB occured.}
\end{figure*}

In Figure 5, we show the log(N2) index \citep[also converted into 12+log(O/H), calibrated using nearby extragalactic HII 
regions, as defined by][]
{pettinipagel}, the H$\alpha$ flux and R23 metallicity as a function of distance in kpc from 
the centre of the galaxy (the centre is taken to be the centre of subaperture 6). In the three 
horizontal panels we provide these values for slit position 1, slit position 2 (for the spiral arm in subapertures 
4 and 5, subaperture 6 is contaminated by a nearby star) and near the GRB location (subaperture 1). 
The log(N2) index, where it was possible to measure, shows an increasing metallicity from the 
southern spiral arm, through the central bulge and into the northern spiral arm. The 
12+log(O/H) value increases from 8.4 in the southern spiral arm to 8.7 in the northern spiral 
arm. This is reinforced by the findings for R23 metallicity, which also shows that the northern 
spiral arm has a higher metallicity than the central bulge and southern spiral arm. We convert 
the R23 metallicity into 12+log(O/H) using the KK04 method in \cite{kewley2008} and break the 
degeneracy between the two solutions using the result for log(N2). The errors on these values can be 
estimated from the errors on the R23 values shown in Figure 5, however there is an uncertainty associated with using the KK04 
method in \cite{kewley2008} which is difficult to quantify. We find the metallicity at 2 kpc from 
the centre of the host to be 8.9 or 8.1 in the southern spiral arm (it was not possible to break the 
degeneracy between the two solutions in this arm), 8.9 in the central bulge, 8.5 in the northern spiral 
arm and 8.5 in slit position 2 at 13.4 kpc from the centre of the galaxy. However, these values for 
metallicity calculated using R23 are likely to be affected by reddening in the host galaxy and the 12+log(O/H) 
metallicity is reliant on breaking the degeneracy of two solutions. Therefore, we base our analysis on the values 
obtained using log(N2) where possible and use the R23 values to corroborate the general result. The values obtained 
for log(N2) are less sensitive to reddening due to the close proximity of the two lines used to calculate this 
value. Taking the solar metallicity to be 8.69 \citep{asplund2004}, we note 
that the southern spiral arm has 0.5 Z$_{\odot}$ and the central bulge and northern spiral 
arms have a value of 1 Z$_{\odot}$. In the southern spiral arm, we can infer a metallicity gradient 
of -0.07 dex kpc$^{-1}$, which is consistent with the Milky Way gradient of -0.09$\pm$0.01 dex kpc$^{-1}$ 
\citep{smartt1997, rolleston2000}. The  H$\alpha$ flux shows a decreasing trend from the southern spiral arm to 
the northern spiral arm. The results from the spiral arm in slit position 2 tend to be in agreement with 
the northern spiral arm in slit position 1. These results show that the southern spiral arm is an actively 
star forming region and this is in direct contrast to the northern spiral arm. The limits at the 
GRB location are provided for reference. The GRB is in the northern spiral arm, on the opposite side of the galaxy to the 
active star formation.

In addition to the emission line properties we measure the 4000 \AA\ break (D4000), which is a 
useful diagnostic for age and metallicity which can even be measured in relatively low signal to 
noise (sub)spectra. Shortward of 4000 \AA\ is the start of stellar photospheric opacity, which
takes into account the mean temperature of the stars. Hotter stars (with shorter lifetimes) have 
more ionised metals in their atmospheres, and hence a lower opacity, than cooler stars. This means 
that an older population of stars will have a higher opacity and, subsequently, a larger 4000 \AA\
break \citep{bruzual,poggianti1997, gorgas1999, kauffmann2003, marcillac2006}. 
\cite{marcillac2006} have shown that D4000 is sensitive to metallicity once the age of the 
population exceeds a few billion years or when it is $>$1.6. D4000 is calculated using the ratio 
between two bands of the continuum, one redwards of the 4000 \AA\ break and the other bluewards.  We use 
the \cite{balogh} definition of the D4000 continua, which is less wide than the original definition 
by \cite{bruzual}, and therefore less affected by dust reddening. The calculated values are plotted 
in Figure 5 and provide a qualitative estimate of the relative ages of the stars as a function of 
position in the galaxy. The estimated error for D4000 (calculated using the RMS of the spectrum and the 
size of the bands) for slit position 1 is $\pm$0.12 and for slit position 2 is $\pm$0.65. 
As expected, it shows that the galactic centre hosts an older population of stars than the spiral 
arms. Interestingly, it also appears that the northern spiral arm hosts an older population of 
stars than the southern spiral arm. This reinforces the evidence of active star formation occuring 
within the southern spiral arm and not in the northern spiral arm. Due to large errors, it was
not possible to calculate D4000 at the GRB location.

Using the approximate metallicity of this galaxy, log(O/H)+12$\sim$8.6 from log(N2), and the 
mass-metallicity relation as measured by \cite{kewley2008} using galaxies in the Sloan Digital Sky 
Survey, we can estimate the mass of the galaxy to be $\sim10^{10}$ M$_{\odot}$. This is consistent
with the value calculated using the near-infrared mass-light ratio.

\section{Discussion}

In previous spatially resolved studies of low redshift GRB host galaxies, it has been determined that LGRBs are associated 
with regions of active star formation and hence provides support for the core collapse supernova progenitor 
theory, for example LGRB 980425 and LGRB 060218 \citep{fynbo2000,wiersema2007,christensen2008}. Additionally, LGRBs at 
higher redshifts tend to occur in the brightest regions of the host galaxy 
\citep{fruchter2006,svensson2010} and relatively small host galaxies \citep{wainwright2007}. GRB 080905A is in direct contrast 
to these results, occuring on the opposite side of a relatively large spiral galaxy to the most active star 
formation region and significantly offset from the centre, so its progenitor is unlikely to be a massive star. The properties 
of this specific region of the host galaxy is in agreement with the findings of \cite{prochaska2006} for typical SGRB environments. 
One of the theoretically predicted progenitors of SGRBs is the merger of a compact binary, for example two
neutron stars or a neutron star and a black hole. Compact binaries are expected to be given a kick velocity during their formation 
which can allow them to travel large distances from their birthplace \citep[][ and references therein]{wang2006}. These 
events are expected to be associated with an older stellar population and offset from the host galaxy, as observed for GRB 080905A.

To summarize, GRB 080905A has short, hard prompt emission with properties expected for a compact binary merger progenitor. 
There was no associated supernova, it appears to be a low density environment and had a low
luminosity. The host galaxy is a spiral galaxy with active star formation, but GRB 080905A occured close to a spiral arm, 
dominated by a relatively old population, and on the opposite side of the galaxy from the spiral arm with most active 
star formation. Additionally, it was offset from the centre of the host galaxy 
by a projected radial distance of 18.5 kpc. Therefore, our observations have shown that GRB 080905A is 
unambigously a short population GRB, whose properties suggest that the progenitor is likely to be a compact 
binary merger. 

\section{Conclusions}

In this paper, we have presented spatially resolved spectroscopy of the host galaxy of the short hard GRB 
080905A, with a T$_{90}$ of 1 s. The prompt emission had an isotropic total energy of $\sim 5\times10^{49}$erg in the energy 
band 15-150 keV. The X-ray and optical afterglows were observed, and the optical afterglow had a magnitude of
$\sim$24 at 8.5 hours after the burst fading to $>$25 at 32 hours.

The host is an almost face on spiral galaxy (inclination $\sim$23$^{\circ}$) with a central bulge and 
at least 2 spiral arms, it is loosely classified as a Sb/c galaxy. The probability that GRB 080905A was chance 
aligned with this galaxy is $<1$\%. The observed redshift of this galaxy 
is $z=0.1218\pm0.0003$, the lowest definite redshift for a typical SGRB thought to originate 
from a compact binary merger. Using spatially resolved 
spectroscopy, we identify a disparity between the two spiral arms, with the southern arm showing a younger 
stellar population and more active star formation than the northern spiral arm. We are unable to be more 
specific as we are using a relative flux callibration, not absolute fluxes, due to the contamination from 
overlying stars and we are not observing the entire host galaxy.

The optical afterglow is observed to be offset from the centre of the galaxy by the projected radial 
distance of 18.5 kpc and occurs in 
the northern region. This offset and the association with an older population in the northern spiral
arm, in addition to the prompt emission properties, shows that GRB 080905A would fit in 
the Type I Gold sample GRB as defined by \cite{zhang2009} with the progenitor being a compact binary 
merger.

\section{Acknowledgements}

AR, KW, AJL and NRT would like to acknowledge funding from the Science and Technology 
Funding Council. PJ acknowledges support by a Marie Curie European Re-integration Grant within the 7th 
European Community Framework Program under contract number PERG03-GA-2008-226653, and 
a Grant of Excellence from the Icelandic Research Fund. The Dark Cosmology Centre is funded by the DNRF. The 
financial support of the British Council and Platform Beta Techniek through the Partnership 
Programme in Science (PPS WS 005) is gratefully acknowledged. We thank A. Van Der Horst, B. Paciesas and 
T. Sakamoto for their help.

This work makes use of data supplied by the UK {\it Swift} Science Data Centre at the University 
of Leicester funded by the Science and Technology 
Funding Council. Based on observations made with the Nordic Optical Telescope, operated
on the island of La Palma jointly by Denmark, Finland, Iceland,
Norway, and Sweden, in the Spanish Observatorio del Roque de los
Muchachos of the Instituto de Astrofisica de Canarias. This research has made use of data obtained 
from the High Energy Astrophysics Science Archive 
Research Center (HEASARC), provided by NASA's Goddard Space Flight Center.

\end{document}